\documentclass[12pt]{iopart}
\usepackage{iopams}
\usepackage{graphicx}
\usepackage[T1]{fontenc}
\usepackage{setstack}
\usepackage{amsfonts}
\usepackage{amssymb}
\usepackage{longtable}

\begin{document}

\title[Clarifications about Lema\^itre-Tolman models of universe]{Some clarifications about Lema\^itre-Tolman models of the Universe used to deal with the dark energy problem}

\author{Marie-No\"elle C\'el\'erier}

\address{Laboratoire Univers et Th\'eories,
Observatoire de Paris, CNRS, Universit\'e Paris Diderot, 5 place Jules Janssen,
92190 Meudon, France}

\ead{marie-noelle.celerier@obspm.fr}

\begin{abstract}

During the past fifteen years, inhomogeneous cosmological models have been put forward to explain the observed dimming of the SNIa luminosity without resorting to dark energy. The simplest models are the spherically symmetric Lema\^itre-Tolman (LT) solutions with a central observer. Their use must be considered as a mere first step towards more sophisticated models. Spherical symmetry is but a mathematical simplification and one must consider spherical symmetric models as exhibiting an energy density smoothed out over angles around us. However, they have been taken at face value by some authors who tried to use them for either irrelevant purposes or to put them to the test as if they were robust models of our Universe. We wish to clarify how these models must be used in cosmology. We first use the results obtained by Iguchi and collaborators to derive the density profiles of the pure growing and decaying mode LT models. We then discuss the relevance of the different test proposals in the light of the interpretation given above. We show that decaying-mode (parabolic) LT models always exhibit an overdensity near their centre and growing-mode (elliptic or hyperbolic) LT models, a void. This is at variance with some statements in the literature. We dismiss all previous proposals merely designed to test the spherical symmetry of the LT models, and we agree that the value of $H_0$ and the measurement of the redshift drift are valid tests of the models. However, we suspect that this last test, which is the best in principle, will be more complicated to implement than usually claimed.

\end{abstract}

\maketitle

\section{Introduction}\label{Intro}

The dimming of the SN Ia luminosity discovered by Riess {\em et al.} (1998) and Perlmutter {\em et al.} (1999) was first interpreted in a standard FLRW cosmology as due to the effect of a cosmological constant or a dark energy component. Since then, a number of alternate solutions have been proposed. Among them, exact inhomogeneous models with no dark energy component were put forward shortly after the release of the first supernova results (Dabrowski and Hendry 1998; Pascual-S\'anchez 1999; C\'el\'erier 2000; Tomita 2000, 2001; Iguchi {\em et al.} 2002). Then, after some neglect, they have started to be reconsidered (R\"as\"anen 2004; C\'el\'erier 2007).

For simplicity sake, most of the first models exhibited spherical symmetry around the observer. This was actually the property used to give the first general analytical derivation of the way dark energy could {\it in principle} be explained away by local inhomogeneities (see Section 2 of C\'el\'erier 2000). However, we wish to dismiss once more a biased interpretation that has been too widely spread since then in the cosmological community. In contrast to what has been advocated by the author of the present paper in the above cited article and by others afterwards, the use of this kind of models {\it should not imply that the observer is physically located at the centre of a spherically symmetric universe.} At the time these models were first proposed, the available cosmological data seemed of course to point to an isotropic universe around us on all cosmological scales. However, since then, more and more precise observations have shown structures on larger and larger scales and we can now conclude that the only nearly isotropic feature in the Universe is the CMB. Close to the last scattering surface, radiative pressure cannot be neglected and dust models such as LT models are no longer relevant.

We therefore propose that one should consider these inhomogeneous spherically symmetric models as exhibiting an energy density smoothed out over angles around us (C\'el\'erier {\em et al.} 2010; Bolejko and Sussman 2011; Bolejko {\em et al.} 2011), i.e. only the radial inhomogeneities are taken into account. This was exemplified by Bolejko and Sussman (2011), who showed, with a realistic Szekeres model, that spherical void structures approximate configurations that can emerge by coarse-graining and averaging a large-scale region of a lumpy universe in which the density distribution is far from spherical.

One should consider that going from the everywhere homogeneous FLRW model to the spherically symmetric LT model merely amounts to switch the use of a solution with three Killing vectors (FLRW) for one with only two Killing vectors (LT). One symmetry (the radial one) is deleted. This was a mere first step to dealing with the cosmological constant problem.

Spherical symmetry around the observer is, of course, a mathematical simplification that is disregarded by more complex models such as Swiss-cheese models with spherical holes (Brouzakis  {\em et al.} 2007; Marra {\em et al.} 2007; Biswas and Notari 2008; Brouzakis {\em et al.} 2008; Marra {\em et al.} 2008),
meatball models (Kainulainen and Marra 2009), Szekeres Swiss-cheese models (Bolejko and C\'el\'erier 2010) etc. Thus, these models must not be taken at face value and some of the proposals designed to test them as if they were exact representations of our Universe, are irrelevant.

Another strange argument against these kinds of models is that they are accused of being {\it non-Copernican}, while the standard spatially homogeneous models are credited to obey the so-called {\it Copernican principle}. However, if one considers an FLRW model from a fully relativistic point of view, i.e. if the location of the observer is considered not only in space but rather {\it in the four-dimensional spacetime}, the model can no more be called Copernican. In reality, the location of the observer {\it in space} is not special, but his {\it temporal} location in the expansion history of the Universe is. In the framework of the dark energy paradigm, this issue is known as the coincidence problem. Therefore, when such an argument is put forward to suggest that FLRW models should be preferred to spherically symmetric ones, this seems rather inconsistent, even from a philosophical point of view.

The spherically symmetric models most commonly used to attempt to explain away dark energy are those of the Lema\^{\i}tre (1933) -- Tolman (1934) (LT) class. These are the ones we discuss in this article.

In Sec. \ref{LT}, we recall the equations defining the LT solutions, mainly to set the notations that are used here. In Sec. \ref{density}, the shape of the mass density profiles of different classes of LT models are studied and some new clarifications are given. Section \ref{obs} is devoted to the discussion of different tests of these models proposed in the literature and, in Sec. \ref{Conclusion}, we present our conclusions.


\section{The Lema\^{\i}tre -- Tolman model}\label{LT}


We recall the equations needed in the following to study the main features of the LT models. We adopt widely used notation defined in, e.g., Krasi\'nski (1997), Pleba\'nski and Krasi\'nski (2006) or Bolejko {\em et al.} (2010).

The LT model is a spherically symmetric non-static solution of Einstein's equations with a dust source. Its metric, in comoving coordinates, synchronous time gauge and in units in which $c=1$, is
\begin{equation}\label{2.1}
{\rm d} s^2 = {\rm d} t^2 - \frac {(R')^2}{1 + 2E(r)}{\rm d} r^2 - R^2(t,r)({\rm
d}\vartheta^2 + \sin^2\vartheta \, {\rm d}\varphi^2),
 \end{equation}
where $E(r)$ is an arbitrary function of integration determining the curvature of space at each $r$ value, a prime denotes a derivative with respect to $r$ and $R(t,r)$ obeys, in units in which $G=1$,
 \begin{equation}\label{2.2}
\dot{R}^2 = 2E + \frac{2M}{R} + \frac{\Lambda}{3} R^2,
 \end{equation}
where a dot denotes a derivative with respect to $t$ and $\Lambda$ is the cosmological constant. Equation (\ref{2.2}) is a first integral of one of Einstein's equations, and $M = M(r)$ is another arbitrary function of integration representing the gravitational mass contained within the comoving spherical shell at a given $r$. The mass density is
 \begin{equation}   \label{2.3}
\kappa \rho = \frac {2M'}{R^2R'}, \qquad {\rm where} \quad \kappa = \frac {8\pi G}
{c^4}.
 \end{equation}

In the following, we assume $\Lambda = 0$. In this case, Eq.~(\ref{2.2}) can be solved explicitly, and the solutions are:

\begin{itemize}

\item when $E < 0$ (elliptic evolution),
\begin{equation}\label{2.4a}
   R(t,r) = \frac{M}{(-2E)}(1 - \cos\eta),
\end{equation}
\begin{equation}\label{2.4b}
   \eta - \sin\eta = \frac {(-2E)^{3/2}}{M} (t - t_B(r)),
 \end{equation}
where $\eta(t,r)$ is a parameter;

\item  when $E = 0$ (parabolic evolution),
 \begin{equation}   \label{2.5}
   R(t,r) = \left[ \frac{9}{2} M (t - t_B(r))^2\right]^{1/3};
 \end{equation}

\item when $E > 0$ (hyperbolic evolution),
\begin{equation}\label{2.6a}
   R(t,r) = \frac{M}{2E} (\cosh\eta - 1),
\end{equation}
\begin{equation}\label{2.6b}
\sinh\eta - \eta = \frac {(2E)^{3/2}}{M} (t - t_B(r)).
 \end{equation}

\end{itemize}

All the formulae presented so far are covariant under arbitrary coordinate
transformations $\tilde{r} = g(r)$, such that $r$ can be chosen at will. This means that one of the three functions $E(r)$, $M(r)$, and $t_B(r)$ can be fixed at one's convenience by an appropriate choice of $g$.


\section{Shape of the mass density profiles of LT models reproducing the $\Lambda$CDM luminosity distance-redshift relation}\label{density}


Most of the density profiles adopted in the literature to construct LT models of a universe with no dark energy that are able to reproduce a significant part of the cosmological data are of the void type (see, e.g., for reviews, R\"as\"anen 2006, C\'el\'erier 2007, Bolejko {\em et al.} 2011, Marra and Notari 2011). It has been proven, however, that a void model is not mandatory to reproduce the main observables of the $\Lambda$CDM model (C\'el\'erier {\em et al.} 2010).

We suspect that the choice of a void is dictated by the reproduction of an accelerated expansion being easier to grasp within such a model. To provide indeed an intuitive understanding of how inhomogeneities can mimic an accelerated expansion, many authors have cited the very pedagogical model proposed by Tomita (2000, 2001). This model is composed of a low-density inner homogeneous region connected at some redshift to an outer homogeneous region of higher density. Both regions decelerate, but, since the void expands faster than the outer region, an observer located inside this void experiences an apparent acceleration when she looks at a source located in the outer region.

There is some confusion in the literature about which models are voids. For instance, the LT example of C\'el\'erier (2000) and the $E=0$ LT case of Iguchi {\em et al.} (2002) have often been presented as void models, which they are not as we show below.

We have seen above that the $r$ coordinate can be chosen at will by fixing the form of one of the three first integrals of Einstein's equations, which are functions of $r$. Here, we make the rather common choice of $M(r) = M_0 r^3$. This choice also has the advantage that it is the one used in the above cited articles, namely C\'el\'erier (2000) and Iguchi {\em et al.} (2002), whose interpretations we intend to specify correctly here.

\subsection{The parabolic (flat) $E=0$ case}\label{parabolic}

We first consider the parabolic (flat) case with $E=0$. With the gauge choice described above, Eq.~(\ref{2.5}) becomes
\begin{equation}\label{2.7}
R = \left(\frac{9 M_0}{2}\right)^{1/3} r (t-t_B)^{2/3}.
\end{equation}
Inserting into Eq.~\ref{2.3}) this expression, its derivative, and the derivative of $M(r)$ with respect to $r$, we obtain
\begin{equation}\label{2.8}
\kappa \rho = \frac{4}{(t-t_B)(3t-3t_B -2r t'_B)}.
\end{equation}
We then calculate the derivative of $\rho$ with respect to $r$
\begin{equation}\label{2.9}
\frac{\kappa \rho'}{4} = \frac{(t-t_B)(8t'_B + 2rt''_B) - 2r t'^2_B}{(t-t_B)^2 (3t-3t_B -2r t'_B)^2}.
\end{equation}
To see whether the observer, with time coordinate $t_0$, is located on a density hill or in a void, we take the limit of the above expression when $r \rightarrow 0$. 
With our gauge choice, we can write $t_B(r) = \tau_1 r + \tau_2 r^2 + {\cal O} (r^3)$, where we have renormalized $t_B$ by translation to be $t_B(r=0) = 0$. Therefore, near the centre, we have $(t_0 - t_B) \rightarrow t_0$, $(3t_0 - 3 t_B - 2r t'_B) \rightarrow 3t_0$ and $2r t_0 t''_B -2r t'^2_B \ll 8 t_0 t'_B \rightarrow 8 t_0 \tau_1$. Thus, we obtain
\begin{equation}\label{2.10}
\kappa \rho'(t=t_0, r \rightarrow 0) \rightarrow \frac{32\tau_1}{9 t_0^3}.
\end{equation}
The same limits applied to Eq.~(\ref{2.8}) give
\begin{equation}\label{2.10a}
\kappa \rho(t=t_0, r \rightarrow 0) \rightarrow \frac{4}{3t_0^2}.
\end{equation}
This implies that, at the observer, the density profile is finite and exhibits a cusp.

Since we can, with no loss of generality, choose $t_0 > 0$, the direction of its slope at the centre, which determines whether the observer is located in a void or on a density hill, is given by the sign of $\tau_1$.

We can first note that, to avoid shell-crossing, $\tau_1 < 0$ (Bolejko {\em et al.} 2010). However, another check can be completed.

The sign of $\tau_1$ is the same as that of $t'_B$ when $r \rightarrow 0$. Iguchi {\em et al.} (2002) studied the two pure classes of $\Lambda = 0$ LT models which reproduce the luminosity distance-redshift relation of the $\Lambda$CDM model, one with $E = 0$ and the other with $t_B = 0$. One can see in their Fig.2 that, for $E = 0$, ${\rm d}t_B / {\rm d} z (z = 0) < 0$.

We now use
\begin{equation}\label{2.11}
t'_B = \frac{{\rm d}t_B}{{\rm d} r} = \frac{{\rm d}t_B}{{\rm d} z} \frac{{\rm d}z}{{\rm d} r}.
\end{equation}
For a radial incoming ray (Bondi 1947),
\begin{equation}\label{2.12}
\frac{{\rm d}z}{{\rm d} r} = \frac{(1 + z) \dot{R}'}{\sqrt{1 + 2E}}.
\end{equation}
In the case $E = 0$,
\begin{equation}\label{2.13}
\frac{{\rm d}z}{{\rm d} r} = (1 + z) \dot{R}',
\end{equation}
with
\begin{equation}\label{2.14}
\dot{R}' = \left(\frac{4 M_0}{81} \right)^{1/3} \frac{3t - 3t_B + rt'_B}{(t - t_B)^{4/3}}.
\end{equation}
At the observer $(t = t_0, r \rightarrow 0)$,
\begin{equation}\label{2.15}
\dot{R}'_0 \rightarrow \left(\frac{4 M_0}{3 t_0} \right)^{1/3} > 0.
\end{equation}
We thus see that the sign of $t'_B$ near the observer, which is that of $\tau_1$, is the same as that of ${\rm d}t_B / {\rm d} z (z = 0)$, i.e., a minus sign.

This implies, therefore, that the observer is located on a density hill and not in a void. As previously noted by C\'el\'erier {\em et al.} (2010), this feature of the density profile only exists on the $t = t_0$ spacelike hypersurface and is thus not directly observable (see however the discussion about the redshift drift in Sec. \ref{obs}).

We therefore wish to stress that the example studied by C\'el\'erier (2000) and the first case considered by Iguchi {\em et al.} (2002) are not void models but overdense ones, in contrast to some incorrect statements in the literature.

\subsection{The simultaneous bang $t_B=0$ case}\label{simbang}

We now consider the simultaneous bang case with $t_B=0$. It is well-known that, with only one degree of freedom left when choosing two of the arbitrary functions of $r$ defining the model, $M$ and $t_B$, the supernova data completely fix the features of the last function $E(r)$. However, we subsequently see that the sign of the derivative of the mass density near the observer is the same in both cases, $E<0$ and $E>0$. We therefore study separately the two cases below, without bothering about whether we know the sign of $E(r)$ at the observer, which is not our purpose here.

\subsubsection{The elliptic $E<0$ case}\label{elliptic}

We first calculate the energy density $\rho(r)$ at the observer. For this, we need the expression of $R'$. Differentiating Eqs.~(\ref{2.4a}) and (\ref{2.4b}) with respect to $r$ and setting $t_B = 0$ and $M= M_0 r^3$, we obtain
\begin{eqnarray}\label{2.21}
R' &=& \frac{M_0 r^2}{(-2E)} \left[3 + \frac{2 r E'}{(-2 E)}\right](1 - \cos\eta) \nonumber \\
&-& \frac{3t}{(-2E)^{1/2}}\left[E' +\frac{(-2 E)}{r}\right] \frac{\sin \eta}{1 - \cos\eta}.
\end{eqnarray}
Inserting this expression and (\ref{2.4a}) in (\ref{2.3}), the energy density is given by
\begin{eqnarray}\label{2.22}
&& \frac{6}{\kappa \rho} = \frac{M_0 r^4}{(-2E)^2}(1 - \cos\eta)^2 \left\{\frac{M_0 r^2}{(-2E)}\left[3 + \frac{2 r E'}{(-2 E)}\right] \right. \nonumber \\
&& \left. \times (1 - \cos\eta) - \frac{3t}{(-2 E)^{1/2}}\left[E' +\frac{(-2 E)}{r}\right] \frac{\sin \eta}{1 - \cos\eta}\right\}.
\end{eqnarray}
We can always, without lack of generality, since $\eta$ is a function of $t$ and $r$, renormalize $t_0$ at the observer such that, at $r=0$, $1 - \cos\eta$ and $\sin \eta$ should be nonzero. In this case, since $E =E_0 r^2 + {\cal O}(r^3)$ and thus $E' = 2E_0 r + {\cal O}(r^2)$, when $r \rightarrow 0$, $\rho$ exhibits a finite value.

We now wish to study the shape of the density function near the observer. To complete the task, we cannot use the same method as in the $E=0$ case, since this would not allow us to draw any definite conclusions. Therefore, we adopt the following reasoning.

From Fig.5 of Iguchi {\em et al.} (2002), we see that ${\rm d}\rho/{\rm d} z > 0$ at the observer. Now,
\begin{equation}\label{2.26}
\rho' = \frac{{\rm d}\rho}{{\rm d} z} \frac{{\rm d} z}{{\rm d} r},
\end{equation}
and for a radial incoming ray,
\begin{equation}\label{2.26a}
\frac{{\rm d}z}{{\rm d} r} = \frac{(1 + z) \dot{R}'}{\sqrt{1 + 2E}}.
\end{equation}
For $r \rightarrow 0$, we have $z \rightarrow 0$, $1 + 2E \rightarrow 1$, and therefore ${\rm d} z/{\rm d} r \rightarrow \dot{R}'_0$. To calculate $\dot{R'}$, we differentiate the expression of $R'$ given by Eq.~(\ref{2.21}) with respect to $t$. This gives
\begin{eqnarray}\label{2.26b}
&& \dot{R}' = \left\{\frac{(-2E)^{1/2}}{r}\left[3 + \frac{2rE'}{(-2E)}\right] - \frac{3}{(-2E)^{1/2}} \right. \nonumber \\
&& \left. \times \left[E' + \frac{(-2E)}{r} \right] \right\} \frac{\sin \eta}{1 - \cos \eta} + \frac{3t(-2E)}{M_0 r^3} \left[E' + \frac{(-2E)}{r} \right] \nonumber \\
&& \times \frac{1}{(1 - \cos \eta)^2}.
\end{eqnarray}
To compute $\dot{R}'_0$ at the observer, it is sufficient to keep the first term in the expansion of $E(r)$ in powers of $r$. We thus obtain
\begin{equation}\label{2.26c}
\dot{R}'_0 = (-2E_0)^{1/2} \frac{\sin \eta_0}{1 - \cos \eta_0}
\end{equation}
Moreover, from Eq.~(\ref{2.4b}), we can write near the centre
\begin{equation}\label{2.26d}
\eta_0 - \sin \eta_0 = \frac{(-2E_0)^{3/2}}{M_0} t_0.
\end{equation}
Since $(-2E_0)$, $t_0$ and $M_0$ are positive, the above equation implies that $\eta_0 > \sin \eta_0$, and therefore $0 < \eta_0 < \pi$, which gives $\sin \eta_0 > 0$. Since we have renormalized $t_0$ so that $1-\cos \eta_0 \neq 0$, we have $1-\cos \eta_0 > 0$. From Eq.~(\ref{2.26c}), we thus obtain $\dot{R}'_0 > 0$ and therefore, $({\rm d} z/{\rm d} r)_0 > 0$. Hence, Eq.~(\ref{2.26}) implies that $\rho'_0$ has the same sign as $({\rm d} \rho/{\rm d} z)_0$, which is strictly positive as $\dot{R}'_0$. In this case, the observer is thus located in a local void.

For an elliptic evolution, we can therefore conclude that the requirement that the bang function should be a constant (always renormalizable to zero), which has been widely assumed, as in, e.g., Yoo {\em et al.} (2008), Zibin (2008) and Zibin {\em et al.} (2008), implies necessarily a void.

This also demonstrates that $\rho'_0$ is non-zero ($\rho$ is non-differentiable at the centre). This was previously noted by Clifton {\em et al.} (2008). In all cases, we claim that this property cannot be used as an argument to dismiss the model, since other features of this kind can be found in nature, e.g., on the surface of the Earth (Bolejko {\em et al.} 2011).

\subsubsection{The hyperbolic $E>0$ case}\label{hyperbolic}

We again first calculate the energy density $\rho(r)$ at the observer. Differentiating with respect to $r$ Eqs.~(\ref{2.6a}) and (\ref{2.6b}) where we have set $t_B = 0$ and $M= M_0 r^3$, we obtain the expression of $R'$
\begin{eqnarray}\label{2.27}
R' &=& \frac{M_0 r^2}{2E} \left(3 - \frac{r E'}{E}\right)(\cosh\eta -1) 
\nonumber \\
&+& \frac{3t}{(2E)^{1/2}}\left(E' -\frac{2 E}{r}\right) \frac{\sinh \eta}{\cosh\eta-1}.
\end{eqnarray}
Inserting this expression and Eq.~(\ref{2.6a}) into Eq.~(\ref{2.3}), the energy density is given by
\begin{eqnarray}\label{2.28}
&& \frac{6}{\kappa \rho} = \frac{M_0 r^4}{(2E)^2}(\cosh\eta-1)^2 \left[\frac{M_0 r^2}{2E}\left(3 - \frac{r E'}{E}\right)(\cosh\eta-1) \right. \nonumber \\
&& \left. + \frac{3t}{(2 E)^{1/2}}\left(E' -\frac{2 E}{r}\right) \frac{\sinh \eta}{\cosh\eta-1}\right].
\end{eqnarray}
As in the previous subsection, we can renormalize $t_0$ at the observer such that, at $\{t=t_0,r=0\}$, $\cosh\eta_0 -1$ and $\sinh \eta_0$ should be nonzero. Therefore, $\rho$ exhibits a finite value there.

We now study the sign of $\rho'_0 = \rho(t=t_0, r \rightarrow 0)$. We have already noted that, according to Fig.5 of Iguchi {\em et al.} (2002), ${\rm d}\rho/{\rm d} z > 0$ near the observer. Using Eq.~(\ref{2.26a}), we have shown above that for $\{t=t_0,r \rightarrow 0\}$, ${\rm d} z/{\rm d} r \rightarrow \dot{R}'_0$. The calculation of $\dot{R}'$ here gives
\begin{eqnarray}\label{2.32}
&& \dot{R}' = \left[ \frac{(2E)^{1/2}}{r} \left(3 - \frac{rE'}{E}\right) + \frac{3}{(2E)^{1/2}} \left(E' - \frac{2E}{r} \right) \right] \nonumber \\
&& \times \frac{ \sinh \eta}{\cosh \eta -1}- \frac{6tE}{M_0 r^3} \left(E' - \frac{2E}{r} \right) \frac{1}{(\cosh \eta - 1)^2}.
\end{eqnarray}
To compute $\dot{R}'_0$, we only keep the first term in the expansion of $E(r)$ in powers of $r$. We thus obtain
\begin{equation}\label{2.33}
\dot{R}'_0 = (2E_0)^{1/2} \frac{\sinh \eta_0}{\cosh \eta_0 - 1}.
\end{equation}
From Eq.~(\ref{2.6b}), we can write
\begin{equation}\label{2.34}
\sinh \eta_0 - \eta_0 = \frac{(2E_0)^{3/2}}{M_0} t_0.
\end{equation}
As in the elliptic case, the right-hand side of this equation is positive, which implies that $\sinh \eta_0 > \eta_0$, hence $\eta_0 >0$ and $\sinh \eta_0 >0$. Since $\cosh \eta_0 -1 > 0$, we obtain $\dot{R}'_0 > 0$. From (\ref{2.26}), we see that $\rho'_0 = ({\rm d} \rho/{\rm d} z)_0 \dot{R}'_0 > 0$. In this case, the observer is thus also located in a local void, which exhibits a cusp at the origin.

As stressed by Krasi\'nski {\em et al.} (2010), since an L--T model is fully specified by two physical functions, a random combination such as $E = 0$ or $t_B = 0$ and an arbitrarily chosen $D_L(z)$, such as the one prevailing in the $\Lambda$CDM picture, may well produce an LT model with unrealistic features.

Since the argument that decaying modes must be avoided to construct a model compatible with the CMB requirements remains disputed (Yoo {\em et al.}  2008; Zibin 2008; Zibin {\em et al.} 2008; C\'el\'erier {\em et al.} 2010), one should consider that intermediate models with both growing and decaying modes can also be used to solve the dark energy problem. This was the result obtained by C\'el\'erier {\em et al.} (2010), where the arbitrary functions determining the LT models capable of reproducing a few features of the $\Lambda$CDM model were computed without any prior assumptions about their form. In these LT models, the shape of the current density profile is not a void, but a hump.

We also recall here that the function $\rho(r)$ considered above {\em is not directly observable}. It exists in the space $t = t_0$ of events simultaneous with our current instant in the cosmological synchronization, i.e. it is in a space-like relation to us. This is also the case for the voids studied by Alnes {\em et al.} (2006), Garc\'ia-Bellido and Haugb\o lle (2008a), and Yoo {\em et al.} (2008).


\section{Test proposals and ruling out claims}\label{obs}


We discuss the main tests and invalidate some claims that have been put forward in the literature. We do not intend to give a review of the subject (this can be found, e.g., in a recent special focus issue of Classical and Quantum Gravity on inhomogeneous cosmological models and backreaction and averaging in cosmology (Anderson and Coley 2011), but only to set the record right about the newest proposals and claims which we find relevant to illustrate our purpose.

To begin with, we wish to stress that all the published tests have been applied to very few particular models of the infinitely larger class of inhomogeneous spherically symmetric models and that, therefore, claims that some test or other rules out the most favoured void model is premature. Moreover, non-void models have seldom been considered but seem to pass some of the tests more successfully.

However, the main drawback of most of these proposals is that they are designed to test the spherical symmetry of the models as if it were an actual physical property of the Universe and not a mere mathematical simplification as we have explained in Sec. \ref{Intro}.

\subsection{The kinematic Sunyaev-Zel'dovich effect}\label{kSZ}

The use of the kinematic Sunyaev-Zel'dovich (kSZ) effect (Sunyaev and Zel'dovich 1972) to test the spherical symmetry of the Universe around the observer was first proposed by Garc\'ia-Bellido and Haugb\o lle (2008b). This proposal involves trying to go beyond the observer's light cone to place constraints on the models. The idea is that, if we are located very near to the centre of a spherical void, we can observe distant sources that are off-centre, which therefore corresponds to observing a large dipole in the CMB spectrum. Such a dipole would manifest itself to us through the kSZ effect. By using available measurements for nine galaxy clusters, the authors claimed that they were able to put constraints on a particular void model.

In recent articles (Zhang and Stebbins 2011, Zibin and Moss 2011, Bull {\em et al.} 2012), the kSZ effect has been once more applied to test different particular void models.

Besides the many loopholes described by Zibin and Moss (2011), the main drawback of the use of the kSZ effect is that, since we are {\it not at the centre of a spherically symmetric universe,} the observer in the distant source is {\it not  off-centre.} Therefore, she is not supposed to observe a large dipole in the CMB and no kSZ effect is therefore observed, in agreement with the measurements. This point was stressed in the discussion of Bull {\em et al.} (2012).

Conversely, if an observer in a distant cluster or free electron cloud were also to use an LT model to analyse her cosmological data, she would find herself apparently ``at the centre'' of a void or a hump. If she were to apply the same reasoning as that given above for the use of the kSZ effect, she would assume that {\it we} are far off-centre in {\it her} spherically symmetric Universe and that {\it we} should see a large dipole in our CMB spectrum. However, this is not the case and the value of the dipole that we measure has allowed us to place stringent bounds on the departures of our location from the centre of a LT model (some tens of Mpc), which are much smaller than those of the remote clusters or free electrons used by Garc\'ia-Bellido and Haugb\o lle (2008b) and Zhang and Stebbins (2011). We therefore see that the above reasoning, which cannot be applied both ways, is unsound.

\subsection{Cosmic parallax}\label{specdis}

The cosmic parallax effect was proposed by Quercellini {\em et al.} (2009) and Quartin and Amendola (2010) to put bounds on the departure of a possible off-centre observer from the symmetry centre of LT models. This effect is based on, in these models, the off-centre observers seeing an anisotropic space. If the expansion is anisotropic, the angular separation between two distant sources varies in time, thereby inducing a cosmic parallax effect, analogous to the well-known stellar parallax, save that this cosmic parallax is caused by a differential cosmic expansion rather than an observer's movement. This effect should be measurable by future space missions such as GAIA or SIM.

However, if this effect can provide an interesting measurement of the general anisotropy of our Universe, it is not adapted to say anything about a universe approximated by LT models. As explained in Sec. \ref{Intro}, these models are only applicable to a central observer, since they are isotropic {\it by construction}.

\subsection{Spectral distortion of the CMB}\label{specdis}

Constraints on void models using the spectral distortion of the CMB have been proposed in the literature (Caldwell and Stebbins 2008; Moss {\em et al.} 2011; Zhang and Stebbins 2011; Zibin and Moss 2011). The philosophy is the same as that underlying the proposal to test LT models with the kSZ effect. A Compton scatterer at a given redshift on the observer past light cone observes an anisotropic CMB that is reflected back to us in the form of spectral distortions, i.e. deviations from a black body spectrum. This {\it y} distortion can be compared to the measurements by FIRAS on board the COBE satellite (Fixsen {\em et al.} 1996), which implies that $y < 15 \times 10 ^{-6}$. Tight constraints were found by Caldwell and Stebbins (2008), who however did not use an LT model but a Hubble bubble one. Other studies completed with pure growing mode LT models found much weaker constraints (Moss {\em et al.} 2011, Zibin and Moss 2011).

Zibin (2011) analysed a specific LT model with pure decaying mode, i.e. a local hump as shown in Sec.~\ref{parabolic}. A generic feature of LT decaying modes is the presence of blueshifts. This means that, in an exact LT universe, the scatterer whose last scattering surface intersects a decaying mode would observe the CMB temperature in the direction of the decaying mode to be greater than the actual temperature at recombination. Therefore, it would see a strongly anisotropic CMB and produce significant Compton spectral distortions. A conclusion of Zibin (2011) is that pure decaying modes that are significant today and wider than 0.02 in redshift are very likely ruled out by this test.

However, this test exhibits the same loophole as those we previously examined, i.e. it is based on the assumption that the Universe is exactly LT. This hypothesis is not needed to validate the use of such models as mere approximations, as we wish to do convince here the reader. Therefore, these three last tests only tell us what we already know from observations, that the Universe is not spatially spherically symmetric around us.

\subsection{The value of $H_0$}\label{Hubble}

In the context of LT models, the most stringent constraint on the present value of the Hubble parameter, $H_0$, comes from the fitting of the CMB power spectrum. This spectrum is the result of, first, the imprint of the primordial perturbations onto the last scattering surface, and second, the geometry of the Universe between this surface and ourselves, which bends the photon trajectory and relates length scales at last scattering to angles measured on our sky. Since perturbation theory applied to an homogeneous FLRW solution is a good approximation for the Universe at last scattering, the simplest solution, usually found in the literature, is an inhomogeneous model where an inner LT region is matched at some redshift to an Einstein-de Sitter (EdS) background. Here, the features of the CMB spectrum are determined, first, by the EdS cosmological parameters, the expansion rate up to last scattering and the relative components of the Universe, second, by the geometry of the model, which differs locally from the standard one owing to the properties of the LT model used. Since we do not have a well-developed perturbation scheme for LT models, the simplified method used at the dawn of inhomogeneous cosmology using these models was as follows: first, calibrate the model by fitting the first peak, then, compute the rest of the spectrum as usual with one's favorite CMB code applied to the EdS model. That it is widely believed that a LT model reproducing the CMB spectrum exhibits a value of $H_0$ too low with respect to the last measured one (Riess {\em et al.} 2011) comes therefore from the following wrong reasoning, based on this simplified scheme.

In the standard $\Lambda$CDM model, if the baryon fraction $\Omega_b$ is fixed, e.g., by nucleosynthesis, there is a strong degeneracy between the fraction of total matter (baryon + dark matter) $\Omega_M$ and the Hubble parameter at the observer $H_0$ for a given size of the angle $\theta$ subtended today on the CMB sky by the sound horizon at recombination. If an LT model is matched to an EdS model at some redshift much lower than that of the last scattering surface, we find that $\Omega_M = 1$ for the CMB. Thus, the size of $\theta$ depends essentially on $H_0$. Now, the ratio of the comoving distance to the recombination $d_C$ to $\theta$ is linked to the scale of the CMB first peak, which is precisely known from the measurements derived from WMAP data. In a given LT model, $d_C$ and $\theta$ are easy to compute. Therefore, the value of $H_0$ depends on $d_C(z) = (1 + z) R[t(z),r(z)]$, i.e., on the two arbitrary functions of the model defining $R(t,r)$. If the background is assumed to be EdS and the bang time function of the LT model is assumed to be constant, then in order to fit a good shape to the CMB power spectrum a low value of the expansion rate is required. This is because the proper shape of the power spectrum demands that $\Omega_M h^2 \approx 0.13$. Therefore, if one assumes  that $\Omega_M =1$, then one gets $h \approx 0.4$. To fit the supernova data, one needs a fluctuation of the expansion rate of amplitude $\delta_H \approx 0.1 - 0.2$ (Enqvist and Mattsson 2007, Bolejko and Wyithe 2009), so this implies that the local expansion rate is low, e.g. $H_0 \approx 45$ km s$^{-1}$ Mpc$^{-1}$ (Moss {\em et al.} 2011).

However, it has been shown by Clifton {\em et al.} (2009) and Bull {\em et al.} (2012) that the local Hubble rate is sensitive to the local bang time and that, by making our region of the Universe younger we can increase $H_0$ up to the measured value.

Another way of increasing the value of $H_0$ was proposed by Biswas {\em et al.} (2010), which is the inclusion of a nonzero overall curvature and a variation in the density profile of the void. Here, the best-fit displayed profiles are compatible with $H_0 = 62.3 \pm 6.3$ km s$^{-1}$ Mpc$^{-1}$, which is the HST value of Sandage {\em et al.} (2006). However, the higher value of $H_0$ measured at the time, equal to $74.2 \pm 3.6$ km s$^{-1}$ Mpc$^{-1}$ (Riess {\em et al.} 2009), could only be obtained at the price of reproducing less accurately the other observations.

Some authors have tried more recently to go deeper into the CMB issue in void models. Clarkson and Regis (2011) demonstrated that void LT models fitting the supernova data and exhibiting a high $H_0$ (= 70 $\pm$ 10 km s$^{-1}$ Mpc$^{-1}$) can reproduce the detailed TT and EE angular power spectra for a flat $\Lambda$CDM model up to l=2500. This is obtained while allowing for the dynamical effect of radiation but ignoring the Integrated Sachs-Wolfe effect. Nadathur and Sarkar (2011) considered a primordial power spectrum that was not assumed to be scale-invariant. They were thus able to simultaneously fit the SNIa data as well as the full CMB power spectrum, while satisfying constraints from local Hubble measurements, primordial nucleosynthesis, and BAO. We note, however, that the value for $H_0$ used here is that obtained by the Hubble Key project $H_0$ = 72 $\pm$ 8 km s$^{-1}$ Mpc$^{-1}$ which is a little smaller than the latest published value of $H_0$ reproduced below.

However, Riess {\em et al.} (2011) claimed that this latest measurement of $H_0 = 73.8 \pm 2.4$ km s$^{-1}$ Mpc$^{-1}$ that they made with the Hubble Space Telescope (HST) and the Wide Field Camera 3 (WFC3) was likely to provide a strong rebuff to giant void models with a central observer. As we have seen above, this is only true for some LT models, but not quite all of them, e.g., those of Clarkson and Regis (2011) and Nadathur and Sarkar (2011).

Finally, Romano (2011), using a local redshift expansion for the luminosity distance and a constraint on the age of the Universe, showed that the parameters defining a general LT model give them enough freedom to enable them to agree with any value of $H_0$. Even if their analysis seems to lack sufficient data to completely constraint the models, and in particular constraints from the CMB, this agrees with the results found by C\'el\'erier {\em et al.} (2010), where LT models reproducing a few features of the $\Lambda$CDM model were computed with no a priori assumption about the form of the density profile. Two separate sets of data were used -- the angular diameter distance together with the redshift-space mass density and the angular diameter distance together with the expansion rate -- both defined on the past null cone as functions of the redshift. The result is not a void but a hump profile and the measured value of $H_0$ is obtained in both cases.

We therefore claim that the value of $H_0$ measured by Riess {\em et al.} (2011) only rules out some LT universe models {\it but not all of them}, be they a void or an overdensity.

\subsection{Redshift drift}\label{redshift}

In our opinion, this is the most robust method for discriminating between different models reproducing the supernova data, as shown for spherically symmetric and homogeneous models by Uzan {\em et al.} (2008).

The redshift drift is the temporal variation in the redshift of distant sources when the observation of the same sources is done at different observer's proper times in an expanding universe. This was first considered by Sandage (1962). In FLRW models, when the expansion of the universe decelerates, all redshifts decrease with time. In models where the expansion is recently accelerating, as in the $\Lambda$CDM model, sources with redshifts $\lesssim 2$ exhibit a positive redshift drift. Uzan {\em et al.} (2008) proposed to use this effect to test the so-called Copernican principle. Other authors have since examined this issue in the framework of the resolution of the dark energy problem with zero-$\Lambda$ LT models (see references in, e.g., the review by Marra and Notari (2011)).

Quartin and Amendola (2010) performed a detailed comparison of three special LT void models with the redshift drift that could be measured in the future by the CODEX experiment, which is a high-precision spectrograph proposed for the European Extremely Large Telescope. They found that, for a mission duration of 15 years and for sources with redshifts $3 < z <4$, the drift could be around $-2 \times 10^{-9}$, which they claim might be measurable by this kind of very high-precision device.

All these authors thus propose that the LT models might be put to the test by such an experiment. This is correct {\it in principle}. However, we recall here that in Sec. \ref{Intro} we stressed that these models are not exact representations of our Universe. Even in the radial direction, the inhomogeneities are somehow smoothed on observable scales. This means that these models {\it cannot be considered as being capable of exactly} reproducing the features of the observed Universe on all scales, and, in particular, at the scales of the local Universe, since the observations show that this local Universe is not spherically symmetric around us as would be true if the models were exact. The very small values calculated for the redshift drift over a timescale of a decade or so might be used to put the different models to the test if these models were exact representations of the Universe in the region of the light cone located between us and the sources. Since this is not the case, either for LT models or FLRW ones, we suspect that the thickness of the lightcone, i.e., the time elapsed between two measurements of the redshifts, needed to properly test these models will be actually longer.

Moreover, some large hump LT models have been shown to be able to exhibit a positive redshift drift as $\Lambda$CDM universes do (Yoo {\em et al.} 2011). This result sets a new challenge for the forthcoming redshift drift experiments, which will have to be more precise than expected to distinguish between inhomogeneous and homogeneous models. The mere sign of the effect will indeed not be enough to draw any strong conclusions, but its amplitude will have to be measured with sufficient accuracy.


\section{Conclusion}\label{Conclusion}

Among the proposals put forward to explain the observed SNIa luminosity without resorting to dark energy, a number of exact inhomogeneous models have been proposed. The simplest and mostly studied ones belong to the Lema\^itre-Tolman class with a central observer. We have once more stressed in this paper that this class must not be considered at face value but as a first step towards developing more sophisticated models. Their use must indeed be viewed as a mere mathematical simplification. The energy density is smoothed out over angles around us (C\'el\'erier {\em et al.} 2010, Bolejko and Sussman 2011, Bolejko {\em et al.} 2011), i.e. only the radial inhomogeneities are taken into account. This is, of course, a very rough approximation which has and will be improved in the future using more sophisticated models to deal with the issue.

Nature does not create objects that fulfill mathematical assumptions with perfect precision. Objects in mechanics or electrodynamics that are described as spherically symmetric have this symmetry only up to some degree of approximation. An ``ideal gas'' in thermodynamics is nearly ideal only at sufficiently low pressure. An ``incompressible fluid'' ..., and so on. However, physicists have always employed simplifying ansatz to be able to complete their calculations, as in the case of exact models that are spherically symmetric on all scales. It is therefore misleading to test these models using methods that are mainly designed to test their simplifying assumptions such as spherical symmetry or the central spatial location of the observer (a violation of the so-called Copernican principle, which, as we have stressed in Sec. \ref{Intro}, is only Copernican in space, not in the full spacetime).

The same remark can be made about FLRW models. They are valid and their robustness has been verified on very large scales where {\it both} the curvature and the density contrasts remain small. However, they no longer provide a valid way to represent our Universe on the scales where these contrasts become too large to allow a perturbative method to converge. We claim that this is the case in the spacetime regions where the supernovae are observed.

We hope that we have succeeded in clarifying these issues and explaining why some statements that can be found in the literature are wrong or vague.

In Sec. \ref{density}, we have shown analytically that the density profile of a LT model reproducing the luminosity distance-redshift relation of the $\Lambda$CDM model is a local hill in the flat case $E=0$ and a local void in the simultaneous bang case, where $t_B = const.$ This strengthens the results obtained by C\'el\'erier {\em et al.} (2010), where the model is a mixture of growing and decaying modes reproducing a few features of the $\Lambda$CDM model and where the density profile is a hump near the observer. Moreover, when compared to the results of Iguchi {\em et al.} (2002), this shows that the energy density functions exhibit the same kinds of profiles for these pure cases in redshift space and on the spacelike hypersurface $t=t_0$.

A few tests applied in the literature to the LT models, and chosen among the most popular ones aiming at ruling out these models, have been discussed in Sec. \ref{obs}.

We have shown that the kSZ effect is irrelevant to put the LT approximation to the test. Since, when using these models, {\it we do not claim we are physically at the centre} of any spherically symmetric universe but that we have merely smoothed out the angular inhomogeneities around us, the observer in the distant source is {\it also not physically off-centre.} Therefore, if she uses the same LT approximation as ourselves, she should not observe a large dipole in the CMB and the lack of kSZ effect seen in the telescopes is explained. The same remark applies to tests based on the cosmic parallax and the spectral distortion of the CMB. If the measurement of the cosmic parallax can provide an interesting evaluation of the general anisotropy of the Universe, it can say nothing about LT approximations. What these three tests can tell us is merely what we already know: that the Universe is not spherically symmetric around us. Since LT models are models of the Universe that are isotropic by construction, hence not exact, it is normal that they should be ruled out by such tests merely designed to test spherical symmetry or anisotropy. In contrast, it would indeed be unusual if they were to pass these tests.

Conversely, the measured value of $H_0$ provides a good test of the models. However, in contrast to what has been advocated by Riess {\em et al.} (2011), and widely believed throughout the community of cosmologists who are not experts in inhomogeneous models, there {\it are} LT models, even void ones, that are compatible with these values. Some have been ruled out, but not all.

The most robust way to test cosmological models would be the measurement of the redshift drift. This could, in principle, help us to distinguish between the models since both the sign and amplitude of the drift depend on the model considered. However, we suspect that this test will need many more years of experiment before it can be said to be conclusive than currently advocated in the literature.

Another criticism of the LT models, which is philosophical and not scientific, cannot be applied to their followers, the Swiss-cheese or meatball models, since they actually comply with the so-called Copernican principle. Moreover, these models tend to represent more physically the observed Universe with its voids and structures. The first Swiss-cheese models made available in the literature use the LT solution to represent the vacuoles spread in the background. They succeed in reproducing the luminosity-distance-redshift relation of the $\Lambda$CDM model, provided that the size of the holes is some hundred Mpc large. Another attempt was made by Bolejko and C\'el\'erier (2010) with Szekeres vacuoles where axial symmetry was used as a first simplifying assumption. In the future, more general Szekeres Swiss-cheese or meatball models should be constructed to deal with the issue.

However, all known exact solutions of the Einstein equations that can be of cosmological use possess some symmetries or quasi-symmetries. The only way to overcome these shortcomings is to obtain a fully operational, exact, and inhomogeneous solution of these equations. This can only be achieved using numerical relativity and we suspect that this will be the new way of dealing with cosmology in the years to come. It will be therefore relevant to confront the models provided by these methods with well-designed cosmological tests.

However, if, at the end of the day, $\Lambda$ appeared to be nonzero or a dark energy component should come into play, such a result, to be robust, should be obtained with methods well-adapted to take the inhomogeneities into account. These could be, either a well-designed averaging scheme, or the use of adapted exact inhomogeneous solutions of General Relativity, probably numerically computed.


\section*{References}
\begin{thebibliography}{99}

\bibitem{Alnes2006}
Alnes, H., Amarzguioui, M. \& Gr\o n, \O. 2006, Phys. Rev. D73, 083519.

\bibitem{AnCo2011}
Anderson, L. \& Coley, A. 2011, CQG 28, 160301.

\bibitem{Bis2008}
Biswas, T. \& Notari, A. 2008, JCAP 0806, 021.

\bibitem{BiNoVa2010}
Biswas, T., Notari, A. \& Valkenburg, W. A. 2010, JCAP 1011, 030.

\bibitem{BoCe2010}
Bolejko, K. \& C\'el\'erier, M.-N. 2010, Phys. Rev. D82, 103510.

\bibitem{BoSu2011}
Bolejko, K. \& Sussman, R. A. 2011, Phys. Lett. B697, 265.

\bibitem{BoWy2009}
Bolejko, K. \&  Wyithe, J. S. B. 2009, JCAP 0902, 020.

\bibitem{BCK2011}
Bolejko, K., C\'el\'erier, M.-N. \& Krasi\'nski, A. 2011, CQG 28, 164002.

\bibitem{BKHC2010}
Bolejko, K., Krasi\'nski, A., Hellaby, C. \& C\'el\'erier M.-N. 2010, {\it Structures in the Universe by Exact Methods: Formation, Evolution, Interactions} (Cambridge:
Cambridge University Press)

\bibitem{Bondi1947}
Bondi, H. 1947, MNRAS 107, 410; reprinted with historical introduction in 1999, GRG 11, 1783.

\bibitem{Brouzakis2007}
Brouzakis, N., Tetradis, N. \& Tzavara, E. 2007, JCAP 0702, 013.

\bibitem{Brouzakis2008}
Brouzakis, N., Tetradis, N. \& Tzavara E. 2008, JCAP 0804, 008.

\bibitem{Bull2012}
Bull, P., Clifton, T. \& Ferreira, P. G. 2012, Phys. Rev. D85, 024002.

\bibitem{CaSte2008}
Caldwell, R. R. \& Stebbins, A. 2008, Phys. Rev. Lett. 100, 191302.

\bibitem{MNC2000}
C\'el\'erier, M.-N. 2000, A\&A 353, 63. 

\bibitem{Cel2007}
C\'el\'erier, M. N. 2007, New Adv. Phys. 1, 29.

\bibitem{CeBK2010}
C\'el\'erier, M. N., Bolejko, K. \& Krasi\'nski, A. 2010, A\&A 518, A21.

\bibitem{Clark2011}
Clarkson, C. \& Regis, M. 2011, JCAP 1102, 013.

\bibitem{Clifton2008}
Clifton, T., Ferreira, P. G. \& Land, K. 2008, Phys. Rev. Lett. 101, 131302.

\bibitem{Clifton2009}
Clifton, T., Ferreira, P. G. \& Zuntz, J. 2009, JCAP 0907, 029.

\bibitem{DabHen1998}
Dabrowski, M. P. \& Hendry, M. A. 1998, ApJ 498, 67.

\bibitem{Enq2007}
Enqvist, K. \& Mattsson, T. 2007, JCAP 0702, 019.

\bibitem{Fixsen1996}
Fixsen, D. J. {\em et al.} 1996, ApJ 473, 576.

\bibitem{GH2008a}
Garc\'ia-Bellido, J. \&  Haugb\o lle, T. 2008a, JCAP 0804, 003.

\bibitem{GH2008b}
Garc\'ia-Bellido, J. \&  Haugb\o lle, T. 2008b, JCAP 0809, 016.

\bibitem{Iguchi2002}
Iguchi, H., Nakamura, T. \& Nakao, K. 2002, Prog. Theor. Phys. 108, 809.

\bibitem{KaMa2009}
Kainulainen, K. \& Marra, V. 2009, Phys. Rev. D80, 127301.

\bibitem{Kras1997}
Krasi\'nski, A. 1997, {\it Inhomogeneous Cosmological Models} (Cambridge: Cambridge
University Press)

\bibitem{KHBC2010}
Krasi\'nski, A., Hellaby, C., Bolejko, K. \& C\'el\'erier, M.-N. 2010, GRG 42, 2453.

\bibitem{Lem1933}
Lema\^{\i}tre, G. 1933, Ann. Soc. Sci. Bruxelles A53, 51; English translation,
with historical comments: 1997, GRG 29, 637.

\bibitem{MaNo2011}
Marra, V. \& Notari, A. 2011, CQG 28, 164004.

\bibitem{Marra2008}
Marra, V., Kolb, E. W. \& Matarrese, S. 2008, Phys. Rev. D77, 023003.

\bibitem{Marra2007}
Marra, V., Kolb, E. W., Matarrese, S. {\em et al.} 2007, Phys. Rev. D76, 123004.

\bibitem{MZS2010}
Moss, A., Zibin, J. P. \& Scott, D. 2011, Phys. Rev. D83, 103515.

\bibitem{Nad2011}
Nadathur, S. \& Sarkar, S. 2011, Phys. Rev. D83, 063506.

\bibitem{Pascual1999}
Pascual-S\'anchez, J. F. 1999, Mod. Phys. Lett. A14, 1539.

\bibitem{Perl1999}
Perlmutter, S. {\em et al.} 1999, ApJ 517, 565.

\bibitem{Ple2006}
Pleba\'nski, J. \& Krasi\'nski, A. 2006, {\it An Introduction to General Relativity
and Cosmology} (Cambridge University Press).

\bibitem{QuAm2010}
Quartin, M. \& Amendola, L. 2010, Phys. Rev. D81, 043522.

\bibitem{Quer2009}
Quercellini, C., Quartin, M. \& Amendola, L. 2009, Phys. Rev. Lett. 102, 151302.

\bibitem{Rasanen2004}
R\"as\"anen, S. 2004, JCAP 0411, 010.

\bibitem{Rasanen2006}
R\"as\"anen, S. 2006, JCAP 0611, 003.

\bibitem{Riess1998}
Riess, A. G. {\em et al.} 1998, AJ 116, 1009.

\bibitem{Riess2009}
Riess, A. G. {\em et al.} 2009, ApJ Supp. 183, 109.

\bibitem{Riess2011}
Riess, A. G. {\em et al.} 2011, ApJ 730, 119 [Erratum-ibid, 732, 129].

\bibitem{Romano2011}
Romano, A. E. 2011, ''Do recent accurate measurements of $H_0$ really rule out void models as alternatives to dark energy?'' arXiv:1105.1864.

\bibitem{Sandage1962}
Sandage, A. 1962, ApJ 136, 319.

\bibitem{San2006}
Sandage, A. {\em et al.} 2006, ApJ 653, 843.

\bibitem{SZ1972}
Sunyaev, R. A. \& Zel'dovich, Y. B. 1972, Comments Astrophys. Space Phys. 4, 173; 1980, MNRAS 190, 413.

\bibitem{Tolm1934}
Tolman, R. C. 1934, Proc. Nat. Acad. Sci. USA 20, 169; reprinted with historical comments: 1997, GRG 29, 931.

\bibitem{Tomita2001}
Tomita, K. 2000, ApJ 529, 382011; 2001, MNRAS 326, 287; 2001 Prog. Theor. Phys. 106, 929.

\bibitem{Uzan2008}
Uzan, J.-P., Clarkson, C. \& Ellis G. F. R. 2008, Phys. Rev. Lett. 100, 191303.

\bibitem{YKN2008}
Yoo, C. M., Kai, T. \& Nakao, K-i. 2008, Prog. Theor. Phys. 120, 937.

\bibitem{Yoo2011}
Yoo, C. M., Kai, T. \& Nakao, K-i 2011, Phys. Rev. D83, 043527.

\bibitem{ZhSt2011}
Zhang, P. \& Stebbins, A. 2011, Phys. Rev. Lett. 107, 041301.

\bibitem{Zibin2008}
Zibin, J. P. 2008, Phys. Rev. D78, 043504.

\bibitem{Zibin2011}
Zibin, J. P. 2011, Phys. Rev. D84, 123508.

\bibitem{ZiMo2011}
Zibin, J. P. \& Moss, A. 2011, CQG 28, 164005.

\bibitem{zib08}
Zibin, J. P., Moss, A. \& Scott, D. 2008, Phys. Rev. Lett. 101, 251303.

\end{thebibliography}
\end{document}